\documentstyle[aps,eepic,amssymb,epic,12pt]{revtex}
\def\<{\langle} \def\>{\rangle}

 \def\Hcal{{\cal H}}  

\def\Tr{\hbox{Tr}}  \def\dim{\hbox{dim}}\def\rank{\hbox{rank}}
\begin{document}
\tightenlines
\title{Improved discrimination of unitaries by entangled probes}
\author{G. Mauro D'Ariano, Paoloplacido Lo Presti, Matteo G. A. Paris}
\address{Quantum Optics $\&$ Information Group \\ Istituto Nazionale per la
Fisica della Materia \\ Universit\`a di Pavia, via Bassi 6, I-27100 
Pavia, ITALIA.} 
\maketitle
\begin{abstract} 
We consider the problem of discriminating among a set of unitaries by
means of measurements performed on the state undergoing the transformation.
We show that use of entangled probes improves the discrimination in the
two following cases: i) for a set of unitaries that are the UIR of a group 
and, ii) for any pair of transformations provided that multiple uses of the 
channel are allowed. 
\end{abstract}
\begin{itemize}\item[]
{\tt E-mail address: paris@unipv.it} \item[]
{\tt URL: www.qubit.it}\end{itemize}
\section{Introduction}
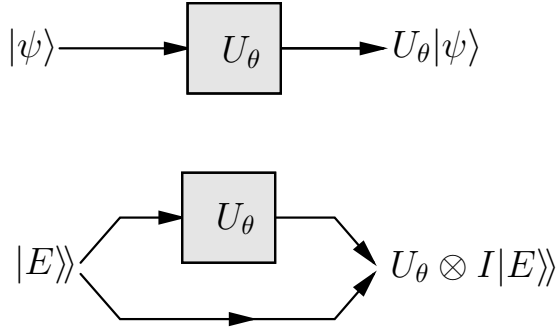
\begin{figure}[hbt]
\begin{center}
\setlength{\unitlength}{0.005mm}
\begin{picture}(10393,8673)(0,-10)
\thicklines
\texture{0 0 0 888888 88000000 0 0 80808 8000000 0 0 888888 88000000 0 0 80808 
        8000000 0 0 888888 88000000 0 0 80808 8000000 0 0 888888 88000000 0 0 80808 }
\path(6847,7404)(9572,7404)
\path(6847,7404)(9572,7404)
\blacken\path(8972.000,7254.000)(9572.000,7404.000)(8972.000,7554.000)(8972.000,7254.000)
\path(972,7404)(4372,7404)
\blacken\path(3772.000,7254.000)(4372.000,7404.000)(3772.000,7554.000)(3772.000,7254.000)
\path(6697,2904)(8272,2904)(9322,1704)
\blacken\path(8814.011,2056.770)(9322.000,1704.000)(9039.784,2254.322)(8814.011,2056.770)
\path(1522,1704)(2572,2904)(4222,2904)
\blacken\path(3622.000,2754.000)(4222.000,2904.000)(3622.000,3054.000)(3622.000,2754.000)
\path(1522,1404)(2572,204)(8272,204)(9322,1404)
\blacken\path(9039.784,853.678)(9322.000,1404.000)(8814.011,1051.230)(9039.784,853.678)
\path(5872,204)(6097,204)
\blacken\path(5497.000,54.000)(6097.000,204.000)(5497.000,354.000)(5497.000,54.000)
\texture{0 0 0 888888 88000000 0 0 80808 8000000 0 0 888888 88000000 0 0 80808 
        8000000 0 0 888888 88000000 0 0 80808 8000000 0 0 888888 88000000 0 0 80808 }
\shade\path(4372,8604)(6847,8604)(6847,6204)
        (4372,6204)(4372,8604)
\path(4372,8604)(6847,8604)(6847,6204)
        (4372,6204)(4372,8604)
\shade\path(4222,4104)(6697,4104)(6697,1704)
        (4222,1704)(4222,4104)
\path(4222,4104)(6697,4104)(6697,1704)
        (4222,1704)(4222,4104)
\put(5272,7029){\makebox(0,0)[lb]{\smash{\large $U_{\theta}$}}}
\put(240,7179){\makebox(0,0)[b]{\smash{\large $|\psi\>$}}}
\put(11000,7179){\makebox(0,0)[b]{\smash{\large $U_{\theta}|\psi\>$}}}
\put(5122,2604){\makebox(0,0)[lb]{\smash{\large $U_{\theta}$}}}
\put(600,1404){\makebox(0,0)[b]{\smash{\large $|E\>\!\>$}}}
\put(12000,1329)
{\makebox(0,0)[b]{\smash{\large $U_\theta\otimes I  |E\>\!\>$}}}
\end{picture}
\end{center}
\caption{The
  parameter $\theta$ is estimated as the result of a unitary
  transformation $|\psi\>\to U_{\theta}|\psi\>$ (up figure).  In this
  scenario the use of a possibly entangled input $|E\>\!\>$ in place of
  $|\psi\>$ is considered, with the unknown transformation
  $U_{\theta}$ acting locally on one Hilbert space only (down
  figure).}
\label{f-scheme}\end{figure}

Entanglement is perhaps the most distinctive ingredient of quantum mechanics.
In the recent years it has been recognised that entanglement is a resource to
improve processing of quantum information and to increase the speed of
computation.  In this paper, we address the use of entanglement as a resource
to improve quantum measurements. In particular, we will deal with measurements
that correspond to the estimation of the parameter $\theta$ labeling a
unitary transformation $U_\theta$ which acts on a system described by the
Hilbert space $\Hcal$.  Usually, the problem is faced by fixing an input state
$|\psi\>\in\Hcal$ that undergoes one of the $U_\theta$'s (Fig.
\ref{f-scheme}), and then applying quantum estimation theory \cite{helstrom}
to look for the POVM which is able to distinguish the possible output states
$U_\theta|\psi\>$ with the minimum error probability $P_E$. In general, this
error probability, or any other chosen figure of merit, will be a function of
the input state $|\psi\>$, and one further optimizes on $|\psi\>$.
\par
Here, we will consider the possibilities offered by the use of a bipartite input 
state $|E\>\!\>\in \Hcal\otimes\Hcal$ instead of the simpler local state $|\psi\>$.  
The transformation $U_\theta$ will act locally on $|E\>\!\>$ thus giving as 
output the state $|\Psi_\theta\>\!\>=U_\theta\otimes I |E\>\!\>$, as depicted 
in Fig. \ref{f-scheme}.  We will show that such novel configuration can do 
better than local measurements in discriminating the unitaries. 
\par 
In Section \ref{s:uni} we focus our attention on the discrimination of unitary 
transformations drawn from a unitary irreducible representation of a group (UIR), 
whereas in Section \ref{s:pair} we will treat the problem of distinguishing among 
two given unitaries. Section \ref{s:out} closes the paper with some concluding
remarks. 
\section{Discrimination among a set of unitary transformations (UIR)}
\label{s:uni}
As a first example consider the problem of discriminating among the
four unitary transformations given by the Pauli matrices
$\{\sigma_i\}$ acting on a qubit. By applying these unitaries to a any
local pure state $|\psi\rangle$, one gets the four non-orthogonal
states $\sigma_j |\psi\rangle$, whereas for a maximally entangled
input state one finds four maximally entangled states which are
orthogonal, and thus exactly distinguishable, at least in principle.
In fact, by adopting the notation $|E\>\!\>\doteq
\sum_{ij}E_{ij}|i\>|j\>$ that puts vectors
$|E\>\!\>\in\Hcal\otimes\Hcal$ into correspondence with operators $E$
on $\Hcal$, a generic maximally entangled input state can be written
as $\frac1{\sqrt d}|U\>\!\>$, with $U$ unitary. Thus, in the Pauli
example the possible outputs are $\frac1{\sqrt2}|\sigma_i\,U\>\!\>$,
and they are orthogonal since $\<\!\<\sigma_i U| \sigma_j U\>\!\>
=\Tr[U^\dag\sigma_i^\dag \sigma_j U]=\delta_{ij}$.  We notice that,
basically, the same kind of configuration has been used for quantum
dense coding. The generalization to a $d$-dimensional system
corresponds to the problem of discriminating the $d^2$ unitary
transformations
\[U(m,n)=\sum_{k=0}^{d-1} e^{2\pi ikm/d}|k\rangle\langle k\oplus n|\;,\]
with $n$ and $m$ ranging in $0\div d-1$, and $\oplus$ denoting
addition modulo $d$. Again, if the input is maximally entangled, we
have orthogonal output states.
\par
Now, suppose we have a set of unitary transformations $\{U_g\}$,
$g\in{\mathbf G}$ that form a (projective) representation of the group
${\mathbf G}$, i. e. $U_gU_h=\omega(g,h) U_{gh}$, where $\omega(g,h)$
is a phase factor satisfying the Jacobi associativity constraints,
namely that $\omega(gh,l)\,\omega(g,h)=\omega(g,hl) \,\omega(h,l)$ and
$\omega(g,g^{-1})=\omega(g,e)=1$, for $g,h,l\in{\mathbf G}$, $e$ being
the identity element. We will consider the case in which such a
representation is irreducible (UIR), i.e. there are no subspaces of $\Hcal$
invariant for the action of all the $U_g$. This was also the case of
the preceding example, with $\{U(m,n)\}$ a UIR of the group
${\mathbb Z}_d\times{\mathbb Z}_d$. Given a UIR, from Schur's lemma it follows
that for each operator $O$ on $\Hcal$ one has
\begin{eqnarray}
\Big[U_gOU^\dag_g\Big]_{\mathbf G}=\mbox{Tr}[O] I\;,
\label{tri}
\end{eqnarray}
where $\big[f(g)\big]_{\mathbf G}$ denotes the group averaging
$\big[f(g)\big]_{\mathbf G}\doteq\sum_{g\in\mathbf G}\mu(g)f(g)$, with
$\mu(g)=\frac{d}{|\mathbf G|}$, $d=\dim(\Hcal)$, and $|{\mathbf G}|$
the cardinality of ${\mathbf G}$. Eq.  (\ref{tri}) can be generalized 
to the continuous case by defining group averaging as $\big[f(g)
\big]_{\mathbf G}\doteq\int_{\mathbf G}
\mu(\mbox{d}g) f(g)$, $\mu(\mbox{d}g) $ being a properly normalized
invariant measure on the group ${\mathbf G}$.
\par
In order to show that entanglement is of help in improving the
discrimination, and to quantify this improvement, we now consider
several state-related parameters.  First of all, as in the first two
examples, one can see that the dimension of the Hilbert space
$\Hcal_{out}$ spanned by the output states is larger for an entangled
input than for factorized states.  In fact, $\dim(\Hcal_{out})$ can be
calculated as the rank of the operator
\begin{equation}
  O= \Big[|\Psi_g\>\!\>\<\!\<\Psi_g|\Big]_{\mathbf G}= \Big[ U_g\otimes I 
  |E\>\!\>\<\!\<E| U_g^\dagger\otimes I \Big]_{\mathbf G}\;,
\label{support}
\end{equation}
where $\Psi_g=U_g E$. By means of Eq. (\ref{tri}) one has $O=I\otimes
\mbox{Tr}_1 [|E\rangle\!\rangle\langle\!\langle E|]=I\otimes (E^\dag
E)^{T}$, so that
\begin{equation} 
\dim(\Hcal_{out})=d\times \rank(E^\dag E)\;,
\end{equation}
i.e. the output space is enlarged by a factor equal to the Schmidt
number \cite{nielsen} of the input state. Indeed, since probing the
operation with a bipartite entangled system gives access to a larger
Hilbert space we have, literally, more room for improvement. In the
following, we refine these concepts, and give conditions under which
and entangled scheme is convenient.
\par The Schmidt number is only a coarse measure of the amount of 
entanglement stored in $|E\>\!\>$, and the dimension of the output space
is only indirectly connected to the distinguishability of the outputs.
A more refined goodness criterion is given by the Holevo's information
$\chi$ of the set of output states, all taken with the same
probability $p(g)=1/|{\mathbf G}|$ (or
$p(\mbox{d}g)=\mu(\mbox{d}g)/\mu({\mathbf G})$ in the continuous
case), this quantity is an upper bound for the accessible information
\cite{nielsen}.  Denoting by $S(\rho)=-\Tr\rho\log\rho$, the von
Neumann entropy of $\rho$, the Holevo's information $\chi$ reads
\begin{eqnarray}
\chi&=&S\left(\frac{1}{\mu({\mathbf G})}\Big[
|\Psi_g\>\!\>\<\!\<\Psi_g|\Big]_{\mathbf G}\right)
-\frac{1}{\mu({\mathbf G})}\Big[S(|\Psi_g\>\!\>\<\!\<\Psi_g|)\Big]_{\mathbf G}=
\nonumber\\
&=& S\left(\frac{1}{\mu({\mathbf G})}I\otimes E^T E^*\right)=\nonumber\\&=&
\frac{d}{\mu({\mathbf G})}\log\mu({\mathbf G})+
\frac{d}{\mu({\mathbf G})}S(E^T E^*)\;,
\end{eqnarray}
and thus the bound is increased by an amount proportional to the
degree of entanglement $S(E^TE^*)$\footnote{\footnotesize $S(E^TE^*)$
  represents the entropy of the partial traces of $|E\>\>$, which
  indeed is the measure of entanglement for pure states.} of the
input state $|E\>\!\>$ (recall that for discrete groups $\mu({\mathbf
  G})=d$).
\par
Facing the problem with a maximum likelihood strategy, the optimal
covariant POVM that discriminates among the $\{|\Psi_g\>\!\>\}$ takes
the form \cite{holevo}
\begin{eqnarray}
 \Pi_g=\mu(g) (U_g\otimes I) P (U_g^\dag\otimes I)\;,
\end{eqnarray}
with $P\ge 0$ a positive operator on ${\cal H}\otimes{\cal H}$
normalized as $\mbox{Tr}_1[P]=I$. By covariance, the likelihood -- i.e.
the probability of getting an outcome $g$ when the state is
$|\Psi_g\>\!\>$ -- is proportional to $ \<\!\<E|P|E\>\!\>\leq d$, where
the bound comes from the normalization condition on $P$, which limits
the largest possible eigenvalue of $P$ below $d$. Again, the
optimality (saturation of the bound) is reached for a maximally
entangled input state, i.e. for $E=d^{-\frac12}U$, with $U$ unitary,
and $P=|U\>\!\>\<\!\<U|$. The optimality of a maximally entangled
input state for the estimation of unitaries in $SU(d)$ has also been
noticed in Ref. \cite{vidal}.
\par
Since the overlap of two states is the only parameter that determines
their distinguishability, we will consider the average overlap
$\Omega(E)$ of all the couples of states in $\{|\Psi_g\>\!\>\}$: the
lower is $\Omega(E)$ the better will be the overall distinguishability. One
has
\begin{eqnarray}
  \Omega(E)&=&\frac1{2\mu({\mathbf G})^2}
\Big[|\<\!\<\Psi_g|\Psi_{g'}\>\!\>|^2\Big]_{{\mathbf G}\times{\mathbf G}}=
\frac1{2\mu({\mathbf G})}\Big[ \<\!\<E|\Psi_{g}\>\!\>
\<\!\<\Psi_{g}|E\>\!\>\Big]_{\mathbf G}=\nonumber\\
&=& \frac1{2\mu({\mathbf G})}\<\!\<E|I\otimes (E^T E^*)|E\>\!\>
= \frac1{2\mu({\mathbf G})} \<\!\<E|EE^\dag E\>\!\>=\nonumber\\
&=& \frac1{2\mu({\mathbf G})} \Tr[(E^\dag E)^2]\;.
\end{eqnarray}
In order to analyze the properties of $\Omega(E)$, we have to briefly
recall the definition of the ``majorization'' relation between
entangled pure states and its physical meaning. Given two states
$|A\>\!\>$ and $|B\>\!\>$ in $\Hcal\otimes\Hcal$, let ${\mathbf
  \lambda}^\downarrow_A$ and ${\mathbf \lambda}^\downarrow_B$ be the
vectors of eigenvalues of $A^\dag A$ and $B^\dag B$ respectively,
sorted in descending order.  We say that $|A\>\!\>\prec|B\>\!\>$ iff
\begin{equation}
\sum_{j=1}^k({\mathbf \lambda}^\downarrow_A)_j
\leq \sum_{j=1}^k({\mathbf \lambda}^\downarrow_B)_j\;,\quad
\mbox{for each $k\leq d$ .}
\end{equation}
The physical meaning of this partial ordering relation has been
clarified in Ref. \cite{nielsen2}: $|A\>\!\>$ can be transformed into
$|B\>\!\>$ by local operations and classical communication if and only
if $|A\>\!\>\prec|B\>\!\>$. Our average overlap $\Omega(E)$ is a so
called ``Schur convex function'' of the eigenvalues of $E^\dag E$,
namely if $|A\>\!\>\prec|B\>\!\>$ then $\Omega(A)\leq\Omega(B)$. Since
any maximally entangled state is majorized by any other state, it is
clear that the minimum overlap is found in correspondence with
$|E\>\!\>$ maximally entangled, and any manipulation of such a state
can only increase $\Omega(E)$, thus reducing the distinguishability, and,
as a consequence, the sensitivity of the measurement. 
\section{Discrimination between two unitary transformations}
\label{s:pair}
Let us suppose that we have to distinguish among two unitaries $U_1$
and $U_2$. Given an input state $|\psi\>$, one optimizes over the possible
measurements, and the minimum error probability
in discriminating $U_1|\psi\>$ and $U_1|\psi\>$ \cite{helstrom} is given by
\begin{eqnarray}
P_E={1\over2}
\left[1-\sqrt{1-|\langle\psi|U_2^\dag U_1|\psi\rangle|^2}\right]\;,\label{PE}
\end{eqnarray}
so that one has to minimize the overlap $|\langle\psi|U_2^\dag
U_1|\psi\rangle|$ with a suitable choice of $|\psi\>$. Chosing as a
basis the eigenvectors $\{|j\>\}$  of $U_2^\dag U_1$, and writing
$|\psi\>=\sum_j \psi_j |j\>$, we define
\begin{equation}
z_\psi\doteq \langle\psi|U_2^\dag U_1|\psi\rangle = \sum_j |\psi_j|^2 
e^{i \gamma_j}\;,\label{Kpoints}
\end{equation}
where $e^{i \gamma_j}$ are the eigenvalues of $U_2^\dag U_1$. The
normalization condition for $|\psi\>$ is $\sum_j|\psi_j|^2=1$, so that
the subset $K(U^\dag_2 U_1)\subset {\mathbb C}$ described by $z_\psi$
for varying $|\psi\>$ is the convex polygon having the points
$e^{i\gamma_j}$ as vertices. The minimum overlap 
\begin{equation}
r(U^\dag_2U_1)\doteq\min_{||\psi||=1}|\langle\psi|U_2^\dag U_1|\psi\rangle|
\end{equation}
is the distance of $K(U^\dag_2U_1)$ from $z=0$. This geometrical
picture indicates in a simple way what is the best one can do in
discriminating $U_1$ and $U_2$: if $K$ contains the origin then the
two unitaries can be exactly discriminated, otherwise one has to find
the point of $K$ nearest to the origin, and the minimum probability of
error is related to its distance from the origin. Once the optimal
point in $K$ is found, the optimal states $\psi$ are those corresponding that
point  through Eq. (\ref{Kpoints}).\par
\begin{figure}[h]
\begin{center}
\setlength{\unitlength}{12mm}
\renewcommand{\dashlinestretch}{30}
\begin{picture}(2.5,2.5)(-1.5,-1.5)
\put(0,0){\circle{2}}\put(-1.1,-0.5){\makebox(0,0){\tiny$\gamma_+$}}
\put(0.6,1){\makebox(0,0){\tiny$\gamma_-$}}
\thinlines
\put(0,-1.25){\vector(0,1){2.5}}\put(-1.25,0){\vector(1,0){2.5}}
\path(0.5,0.86)(-0.86,0.5)(-0.94,-0.342)(0.5,0.86)
\texture{8101010 10000000 444444 44000000 11101 11000000 444444 44000000 
        101010 10000000 444444 44000000 10101 1000000 444444 44000000 
        101010 10000000 444444 44000000 11101 11000000 444444 44000000 
        101010 10000000 444444 44000000 10101 1000000 444444 44000000 }
\shade\path(0.5,0.86)(-0.86,0.5)(-0.94,-0.342)(0.5,0.86)
\thicklines
\drawline(0,0)(-0.22,0.26)
\put(-.1,0.15){\tiny$r$}
\end{picture}
\caption{$r$ is the minimum distance between the origin and the
polygon $K$}\label{figurilla}
\end{center}
\end{figure}
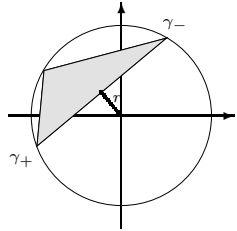
If $\Delta(U_2^\dag U_1)$ is the angular spread of the eigenvalues of
$U_2^\dag U_1$ (referring to Fig. \ref{figurilla}, it is
$\Delta=\gamma_+-\gamma_-$), from Eq. (\ref{PE}) for $\Delta< \pi$
one has
\begin{equation}
P_{E}=\frac12-\frac12\sqrt{1-\cos^4\frac{\Delta}2}\;,
\label{PEdelta}
\end{equation}
whereas for $\Delta\ge \pi$ one has $P_E=0$ and the discrimination is exact.
\par
Given $U_1$ and $U_2$ non exactly discriminable, one is interested in
understanding wheter or not an entangled input state could be of some
use.  The answer is negative, in fact using entanglement translates
the problem into the one of distinguishing between $U_1\otimes I$ and
$U_2\otimes I$, thus one has to analyze of the polygon $K(U_2^\dag
U_1\otimes I)$. Since $U_2^\dag U_1\otimes I$ has the same eigenvalues
as $U_2^\dag U_1$, the polygons $K(U_2^\dag U_1\otimes I)$ and
$K(U_2^\dag U_1)$ are exactly the same, so that they lead to the same
minimum probability of errror.
\par
\begin{figure}[hbt]
\begin{center}
\setlength{\unitlength}{0.005mm}
\begin{picture}(9540,11898)(0,-10)
\thicklines
\texture{0 0 0 888888 88000000 0 0 80808 8000000 0 0 888888 88000000 0 0 80808 
        8000000 0 0 888888 88000000 0 0 80808 8000000 0 0 888888 88000000 0 0 80808 }
\shade\path(3525,11829)(6000,11829)(6000,9429)
        (3525,9429)(3525,11829)
\path(3525,11829)(6000,11829)(6000,9429)
        (3525,9429)(3525,11829)
\put(4275,10404){\makebox(0,0)[lb]{\smash{\large$U$}}}
\shade\path(3600,2454)(6075,2454)(6075,54)
        (3600,54)(3600,2454)
\path(3600,2454)(6075,2454)(6075,54)
        (3600,54)(3600,2454)
\put(4350,1029){\makebox(0,0)[lb]{\smash{\large$U$}}}
\shade\path(3525,9129)(6000,9129)(6000,6729)
        (3525,6729)(3525,9129)
\path(3525,9129)(6000,9129)(6000,6729)
        (3525,6729)(3525,9129)
\path(2025,7929)(3525,7929)
\blacken\path(2925.000,7779.000)(3525.000,7929.000)(2925.000,8079.000)(2925.000,7779.000)
\path(1125,6129)(2025,7929)
\path(1125,6804)(2025,10629)
\path(1133,6802)(2033,10627)
\path(6000,7929)(7425,7929)(8325,6129)
\blacken\path(7922.508,6598.574)(8325.000,6129.000)(8190.836,6732.738)(7922.508,6598.574)
\path(6000,10629)(7425,10629)(8325,6729)
\blacken\path(8043.926,7279.906)(8325.000,6729.000)(8336.243,7347.364)(8043.926,7279.906)
\path(2025,10554)(3525,10554)
\blacken\path(2925.000,10404.000)(3525.000,10554.000)(2925.000,10704.000)(2925.000,10404.000)
\path(1117,4996)(2017,1171)
\path(1950,1179)(3450,1179)
\blacken\path(2850.000,1029.000)(3450.000,1179.000)(2850.000,1329.000)(2850.000,1029.000)
\path(6150,1179)(7575,1179)(8475,5079)
\blacken\path(8486.243,4460.636)(8475.000,5079.000)(8193.926,4528.094)(8486.243,4460.636)
\path(1125,5529)(3525,5529)
\path(5925,5529)(8325,5529)
\blacken\path(7725.000,5379.000)(8325.000,5529.000)(7725.000,5679.000)(7725.000,5379.000)
\put(4275,7629){\makebox(0,0)[lb]{\smash{\large$U$}}}
\put(3850,5229){\makebox(0,0)[lb]{\smash{\huge$\cdots$}}}
\end{picture}
\end{center}
\caption{When distinguishing between two unitaries $U=U_{1,2}$ it is possible
  to achieve perfect discrimination even for nonorthogonal $U_1$ and
  $U_2$ for sufficiently large number $N$ of copies of the unitary
  transformation, using a $N$-partite entangled state as in figure (see text).}
\label{N-times}\end{figure}
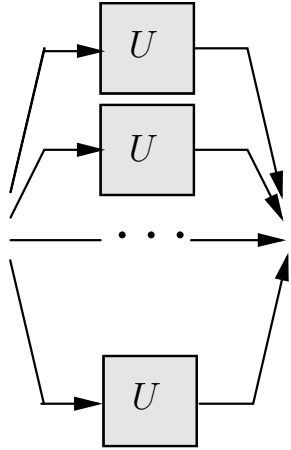

The situation changes dramatically if $N$ copies of the
unitary transformation are used, as depicted in
Fig. \ref{N-times}: here one has to compare the
``performance'' of $K(U_2^\dag U_1)$ to the one of $K((U_2^\dag
U_1)^{\otimes N})$.
Since $\Delta((U_2^\dag U_1)^{\otimes N})=\min\{N\times\Delta(U_2^\dag
U_1),2\pi\}$, it is clear that there will be an $\bar N$ such that
$U_1^{\otimes N}$ and $U_2^{\otimes N}$ will be exactly discriminable.
This same result has been demonstrated in Ref. \cite{acin} starting from a
different approach. 
\section{Conclusions} \label{s:out}
We have shown that the use of entangled states as a probe provides 
an effective scheme to discriminate among a set of unitary transformations.
We have analyzed the discrimination of a set of unitaries 
being the UIR of a group, showing that entanglement is always useful.
We have also considered the discrimation between two generic transformations, 
where it is possible to achieve perfect discrimination even for nonorthogonal 
$U_1$ and $U_2$ for sufficiently large number $N$ of copies of the unitary
transformation, if a $N$-partite entangled state is available.
The present results for the discrimination of a {\em discrete} 
set of unitaries can be generalized to the continuos case \cite{prl}, 
{\em i.e.} to the estimation of parameters. In this case
entanglement improves the performances of the measurment scheme
also in presence of losses.

\end{document}